\documentclass[12pt]{article}

\usepackage{axodraw}

\newcommand{\be}{\begin{equation}}
\newcommand{\ee}{\end{equation}}
\newcommand{\ben}{\begin{equation*}}
\newcommand{\een}{\end{equation*}}

\newcommand{\ra}{\rangle}
\begin{document}
\begin{titlepage}
%\hskip 11cm \vbox{ \hbox{Budker INP 2006-??}} \vskip
%0.3cm
\begin{center}
{\bf Proof of the multi-Regge form of QCD amplitudes
\\with gluon exchanges in the NLA$^{~\ast}$}
\end{center}
\vskip 0.5cm \centerline{V.S.~Fadin$^{a\,\dag}$,
R.~Fiore $^{b\,\ddag}$, M.G.~Kozlov$^{a\,\dag\dag}$,
A.V.~Reznichenko$^{a\,\ddag\dag}$} \vskip .3cm
\begin{center}
\vskip .3cm \centerline{\sl $^{a}$ Budker Institute of
Nuclear Physics, 630090 Novosibirsk, Russia}
\centerline{\sl Novosibirsk State University,
630090 Novosibirsk, Russia} \centerline{\sl $^{b}$
Dipartimento di Fisica, Universit\`a della Calabria,}
\centerline{\sl Istituto Nazionale di Fisica
Nucleare, Gruppo collegato di Cosenza,} \centerline{\sl
Arcavacata di Rende, I-87036 Cosenza, Italy} \vskip 1cm
\end{center}
\vskip 1cm

%\thanks{Work supported by the Russian Fund of Basic
%Researches, grant 04-02-16685}
%V.S. Fadin
%\thanks {\it Humboldt Preistr\"ager}}
%$^{(\dagger)}$
%\\ Institute of Nuclear
%Physics and Novosibirsk State University,\\
%       630090 Novosibirsk, Russia%

\begin{abstract}

The multi--Regge form of QCD amplitudes with gluon
exchanges is proved in the next-to-leading approximation.
The proof is  based on the bootstrap relations, which are
required for the compatibility of this form with the
s-channel unitarity. We show  that  the fulfillment of
all these  relations ensures the Reggeized form of energy
dependent radiative corrections order by order in
perturbation theory. Then we prove that all  these
relations are fulfilled  if  several bootstrap conditions
on the Reggeon vertices and trajectory hold true. Now all
these conditions are checked and proved to be satisfied.

\end{abstract}
%\vskip .5cm
\vfill \hrule \vskip.3cm \noindent $^{\ast}${\it Work
supported in part by INTAS, in part by the Russian
Fund of Basic Researches and in part by Ministero
Italiano dell'Istruzione, dell'Universit\`a e della
Ricerca.} \vfill $
\begin{array}{ll} ^{\dag}\mbox{{\it e-mail address:}} &
\mbox{FADIN@INP.NSK.SU}\\
^{\ddag}\mbox{{\it e-mail address:}} &
\mbox{FIORE@CS.INFN.IT}\\
^{\dag\dag}\mbox{{\it e-mail address:}} &
\mbox{M.G.KOZLOV@INP.NSK.SU}\\
^{\ddag\dag}\mbox{{\it e-mail address:}} &
\mbox{A.V.REZNICHENKO@INP.NSK.SU}\\
\end{array}
$
\end{titlepage}
\vfill
\eject
\section{Introduction}

Reggeization of gluons as well as
quarks~\cite{GST}-\cite{BLF} is one of remarkable
properties of Quantum Chromodynamics (QCD). The gluon
Reggeization is especially important since cross sections
non vanishing in the high energy limit  are related to
gluon exchanges in cross channels.  A primary Reggeon in
QCD turns out to be the Reggeized gluon.

The gluon Reggeization  gives the most common basis for
the description of high energy processes.  In particular,
the famous BFKL equation \cite{BFKL} was derived
supposing the Reggeization. The most general approach to
the unitarization problem is the reformulation   of QCD
in terms of  a gauge-invariant effective field theory for
the Reggeized gluon interactions \cite{Lipatov:1995pn}.

Let us emphasize that we use the term ``Reggeization" in a
much stronger sense than the existence of the Reggeon with
gluon quantum numbers and trajectory $j(t)=1+\omega(t)$
with $\omega(0)=0$. We use it as the statement that
contributions of solely this Reggeon determine the high
energy behaviour of all QCD amplitudes in the multi-Regge
kinematics (including, of course, the Regge kinematics as
a particular case).

The gluon Reggeization was proved \cite{BLF} in the
leading logarithmic approximation (LLA), i.e. in the case
of  summation of the terms $~(\alpha _S\ln s)^n$ in
cross-sections of processes at energy $\sqrt s$ in the
c.m.s., but till now remains a hypothesis in the next-to-leading
 approximation (NLA), when the terms $~\alpha
_S(\alpha _S\ln s)^n$ are also kept. Now the BFKL
approach, based on the gluon Reggeization,  is
intensively developed in the NLA; in particular, the BFKL
kernel is known now both for forward \cite{FL98} and
non-forward \cite{FF05} scattering.  Moreover, some
effective Reggeon-particle vertices, which can be used for
the development of the next-to-next-to-leading
approximation, are also calculated (see, for instance,
Ref.~\cite{Antonov:2004hh}). Meanwhile, there is the
statement \cite{Kucs} that the Regge form of QCD
amplitudes is violated at the three-loop level in the
next-to-leading order (NLO), that means, in our
terminology,  absence of the gluon Reggeization in the
NLA. It makes extremely important the problem of proving
or rejecting  the Reggeization hypothesis in this
approximation.

A possible way of solution of this problem was outlined
in Ref.~\cite{V.F.02}. It is based on  the ``bootstrap"
relations, which are required for the compatibility of the
gluon Reggeization with the $s$-channel unitarity. In
this paper we present the proof of the gluon Reggeization
in the NLA which is  obtained in this way. First we show
that the fulfillment of the bootstrap relations
guarantees the multi--Regge form of QCD amplitudes. Then
we demonstrate that an infinite set of these bootstrap
relations are fulfilled if several conditions imposed on
the Reggeon vertices and the trajectory (bootstrap
conditions) hold true. Now all these conditions  are
proved to be satisfied, and this means that the gluon
Reggeization is true, contrary to the statement of
Ref.~\cite{Kucs}.

To be definite, we have to say that our consideration is
limited by QCD perturbation theory. Our particles are
actually partons --- quark and gluons. Moreover, we
confine ourselves in the framework of the NLA, and all
our assertions and equations given below must be  taken
with this accuracy.

\section{Multi-Regge form of QCD amplitudes}

Objects of our investigation are QCD amplitudes in the
multi-Regge kinematics (MRK). We call MRK  the kinematics
where all particles have {limited (not growing with $s$)
transverse momenta} and are combined into jets with
{limited invariant mass of each jet and large} (growing
with $s$) {invariant masses of any pair of the jets}. The
MRK gives {dominant contributions to cross sections} of
QCD processes at high energy $\sqrt s$. At that in the
LLA  only gluons are produced and each jet is actually a
gluon. In the NLA one of  jets can contain a couple of
particles (two gluons or quark-antiquark pair). Such
kinematics is called also {quasi multi-Regge kinematics
(QMRK)}. We use the notion of jets and extend the notion
of MRK, so that it includes the QMRK,  in order to unify
considerations.

Let us consider the amplitude ${\cal A}_{2\rightarrow n+2}$
of the process $A+B\rightarrow A'+J_1+\ldots+J_n+B'$ in the
MRK. We will use light-cone momenta $n_1$ and $n_2$, with
$n_1^2=n_2^2=0,\;\;(n_1n_2)=1$,  and denote
$(pn_2)\equiv p^{+},\;\; (pn_1)\equiv
 p^{-}$. We assume that initial momenta
$p_A$ and $p_{B}$ have predominant components $p_A^+$ and
$p_B^-$.  For generality it is not assumed that the
components $p_{A\bot},\;\;p_{B\bot}$ transverse to
the $(n_1, n_2)$ plane are zero. Moreover,  $A$ and
$B$, as well as $A^{\prime}$ and $B^{\prime}$, can
represent jets. We suppose that rapidities of final jets
$J_i$ with momenta $k_i$
$y_i=\frac{1}{2}\ln\left(k^+_i/k^-_i\right)$ decrease
with $i$: $y_0> y_1>\dots >y_n> y_{n+1}$; as for $y_0$ and
$y_{n+1}$, it is convenient to define them as $y_0=
y_A\equiv\ln\left(\sqrt 2 p^+_A/|q_{1\bot}|\right)$ and
$y_{n+1}=y_B\equiv \ln\left(|q_{(n+1)\bot}|/\sqrt 2
p^-_{B}\right)$. Notice that $q_i$ indicate the Reggeon momenta
and $q_1=p_{A^{\prime}}-p_{A}\equiv q_A$, $q_{n+1}=p_{B}-p_{
B^{\prime}}\equiv q_B$.

Our aim is to prove that in the NLA the real part of the
amplitude ${\cal A}_{2\rightarrow n+2}$ has the
multi-Regge  form
\begin{equation}
{\cal A}^{R}_{2\rightarrow n+2}=\bar{\Gamma}^{{R}_1}_{
A^{\prime}A} \left( \prod_{i=1}^n
 \frac{e^{\omega(q_i)(y_{i-1}-y_i)}}{q^2_{i\perp}}\gamma^{J_i}_{{R}_i
 {R}_{i+1}}
 \right) \frac{e^{\omega(q_{n+1})(y_{n}-y_{n+1})}}{q^2_{(n+1)\perp}}
 \Gamma^{{R}_{n+1}}
_{ B^{\prime} B}. \label{A 2-2+n}
\end{equation}
Here $\omega(q)$ is called gluon Regge trajectory,
%
%\vspace{0.5 cm} \hspace{-0.7cm} \epsfig{file=fig.eps}
%\noindent
$\;\;\Gamma^{{R}} _{ B^{\prime} B}$ and
$\bar{\Gamma}^{{R}}_{ A^{\prime}A}$ are the scattering
vertices, i.e. the effective vertices for $B \rightarrow
B^{\prime}$  and $A \rightarrow A^{\prime}$ transitions
due to interaction with Reggeized gluons ${R}$; $\;
\gamma ^{J_{i}}_{{R}_i {R}_{i+1}}$ are the production
vertices, i.e. the effective vertices for production of
jets $J_i$ with momenta $k_i=q_{i+1}-q_i$ in
$R_{i+1}\rightarrow {R}_i$ transition of Reggeons with
momenta $q_{i+1}$ and $q_{i}$. We use for particles and
Reggeons notations which accumulate all their quantum
numbers. All Reggeon vertices, as well as the gluon
trajectory, are known now with the required accuracy (see
Ref.~\cite{VF03} and references therein;  the scattering
vertices $\;\;\Gamma^{{R}} _{ B^{\prime} B}$ and
$\bar{\Gamma}^{{R}}_{ A^{\prime}A}$ in Eq.~(\ref{A
2-2+n}) differ by the factors $2p^-_B$ and $2p^+_A$
correspondingly from the analogous quantities used
there).

Remind that as compared with ordinary particles Reggeons
possess an additional quantum number, the signature, which
is negative for the Reggeized gluon. In each order of
perturbation theory amplitudes with negative signature do
dominate, owing to the cancellation of leading
logarithmic terms in amplitudes with  positive signature
which become pure imaginary in the leading order for them
(which coincides  with the next-to-leading for negative
signature). We emphasize that only the real parts of the
amplitudes have the representation (\ref{A 2-2+n}). Only
these parts have such a simple form, and only these parts
are given by the Reggeized gluon contributions. As for
imaginary parts, they come into amplitudes both from the
parts with positive and negative signatures. They can be
calculated using the unitarity relations and the
amplitudes (\ref{A 2-2+n}). It is well known from the
BFKL equation for the Pomeron exchange that they are
complicated even for elastic amplitudes.

Let us show that the amplitudes (\ref{A 2-2+n}) have
negative signatures in all $q_i$--channels.  In order to
construct amplitudes with definite signatures one needs
to perform the ``signaturization". In general the
signaturization is not a simple task. It requires
partial-wave decomposition of amplitudes in
cross-channels with subsequent symmetrization
(anti-symmetrization) in ``scattering angles" and
analytical continuation into the $s$--channel. The
procedure is relatively simple only in the case of
elastic scattering of spin-zero particles. At that,
generally speaking, even in this case the amplitudes with
definite signatures cannot be expressed in terms of
physical amplitudes related by crossing. Fortunately, at
high energy the signaturization can be easily done not
only for elastic, but in the MRK also for inelastic
amplitudes, for particles with spin as well as for the
spin-zero ones. The signaturization (as well as crossing
relations) is naturally formulated for ``truncated"
amplitudes, i.e. for amplitudes  with omitted wave
functions (polarization vectors and Dirac spinors). The
crucial points are that in the MRK all energy invariants
$s_{i,j}=(k_i+k_j)^2$ are large and that they are
determined only by the longitudinal components of momenta
($s_{i,j}=2k_i^+k_j^-, \;\; i<j$). Due to largeness of
$s_{i,j}$  signaturization in the $q_l$--channel means
symmetrization (anti-symmetrization) with respect to the
substitution $s_{i,j}\leftrightarrow -s_{i,j}, \;\;
i<l\leq j$. Since $s_{i,j}$ are determined by
longitudinal components, this substitution is equivalent
to the replacement $k^{\pm}_i\leftrightarrow
-k^{\pm}_i,\;\;i<l,\;\; p_A^{\pm}\leftrightarrow
-p_A^{\pm}$  (or, equivalently, $k^{\pm}_j\leftrightarrow
-k^{\pm}_j,\;\;j\geq l,\;\; p_B^{\pm}\leftrightarrow
-p_B^{\pm}$) in the truncated amplitudes  without change of
transverse components. Note that such substitution does
not violate the total momentum conservation due to the strong
ordering of the longitudinal components. At that all
particles remain on their mass shell, so that the
substitution is equivalent to the transition into the
cross-channel. Note that the limitation by the real parts in
the form (\ref{A 2-2+n}) means that the Regge factors
remain unchanged under the crossing.

In order to understand the behaviour of the amplitudes
(\ref{A 2-2+n}) under the signaturization   it is
convenient to take the  production vertices $\; \gamma
^{J_{i}}_{{R}_i {R}_{i+1}}$  in the physical light--cone
gauges with gauge--fixing vectors $n_2$ or $n_1$. Then it
becomes evident that these vertices do not depend on
longitudinal components of momenta and remain unchanged
under the crossing, whereas the scattering vertices
entering in the form (\ref{A 2-2+n}) change their signs
due to the discussed factors $p_A^+$ and $p_B^-$.  It
ensures negative signature in all $q_i$--channels.

The factorized form of QCD amplitudes in the {MRK} was
proved at the Born level using the $t$--channel unitarity
and analyticity \cite{BFKL}.  Their Reggeization  was
first derived in the LLA on the basis of the direct
calculations at the three-loop level for elastic
amplitudes and the one-loop level for one-gluon
production amplitudes. Later {it was proved \cite{BLF} in
the LLA for all amplitudes at arbitrary number of loops}
with the help of {bootstrap relations}. {At NLA the
Reggeization remained a hypothesis till now.}

{The hypothesis is extremely powerful} since an infinite
number of amplitudes is expressed in terms of the gluon
Regge trajectory and several Reggeon vertices.

\section{{\bf  Bootstrap relations}}

The proof of the form (\ref{A 2-2+n}) is based on   the
{$s$-channel unitarity}, which provides us with the
discontinuities $\mathrm{disc}_{s_{i,j}}$ of the signaturized
amplitude ${\cal A}^{\cal S}_{2\rightarrow n+2}$ in the
channels $s_{i,j}=(k_i+k_j)^2$. Note that generally
speaking these discontinuities  are not pure imaginary in
the NLA, since  a discontinuity in one of the channels
can have, in turn, a discontinuity in another channel.
But it is clear that these double discontinuities are
sub-sub-leading, so that  we will neglect them in the
following.

For elastic amplitudes the connection of real parts of
the amplitudes and their discontinuities is well known.
Unfortunately, it is quite not so for {inelastic
amplitudes}. Analytical properties of the production
amplitudes are very complicated even in the MRK
\cite{Bartels_1}.  But, fortunately,  if in the MRK we
confine ourselves to the NLA, these properties are
greatly simplified and allow us \cite{V.F.02} to express
partial derivatives ${\partial }/{\partial y_{j}}$ of the
amplitudes, considered as  functions of rapidities $y_j
\;\; (j=0, \dots, n+1)$ and transverse momenta, in terms
of the discontinuities of the signaturized amplitudes:
\be \frac{1}{-\pi
i}\left(\sum_{l=j+1}^{n+1}\mathrm{disc}_{s_{j,l}}
-\sum_{l=0}^{j-1}\mathrm{disc}_{s_{l,j}}\right){\cal
A}^{\cal S}_{2\rightarrow n+2}/(p_A^+p_B^-) \,
=\,\frac{\partial }{\partial y_{j}}\; {\cal A}^{\cal
S}_{2\rightarrow n+2}(y_i)/(p_A^+p_B^-)\,. \label{deriv}
\ee  Note that taking the sum of the equations
(\ref{deriv}) over $j$ from $0$ to $n+1$ it is easy to
see from Eq.~(\ref{deriv}) that  ${\cal A}^{\cal
S}_{2\rightarrow n+2}$ depends only on differences of the
rapidities $y_i$, as it must be. The division by
$(p_A^+p_B^-)$  is performed in Eq.~(\ref{deriv}) in
order to differentiate the rapidity dependence  of
radiative corrections only.

Equalities (\ref{deriv})  can be easily proved using
the Steinmann theorem \cite{S:1960}, or, more
definitely, the statement \cite{Bartels_1} that the
amplitudes can be presented as a sum of contributions
corresponding to various sets of the $n+1$ non-overlapping
channels $s_{i_k,j_k}\,,\;i_k<j_k, \;\; k=1,\dots,n+1$;
at that each of the contributions can be written as a
signaturized  series in logarithms of the energy variables
$s_{i_k,j_k}$ with coefficients which are a real function of
transverse momenta.  Remind that two channels
$s_{i_1,j_1}$ and $s_{i_2,j_2}$ are called overlapping if
either $i_1<i_2\leq j_1< j_2$, or $i_2<i_1\leq j_2< j_1$.
Since scattering amplitudes enter in the relations
(\ref{deriv})  linearly and uniformly, it is sufficient
to prove these relations  separately for the contribution
of one of the sets. Now two  observations are
important: first, we need not to consider the
coefficients depending on transverse momenta  neither
calculating  the discontinuities, nor calculating
the derivatives over $y_j$ in Eq.~(\ref{deriv}); and second,
the energy variables $s_{i_k,j_k}$  entering in each set  are
independent, i.e. there are no relations between
the differences $y_{i_k}-y_{j_k}$  for non-overlapping
channels $s_{i_k,j_k}$; this means, in particular, that we
need to consider only leading and next-to leading orders
in logarithms of these variables.

Therefore,  it is sufficient to prove the equalities
(\ref{deriv})  with the NLO  accuracy for the symmetrized
products
\begin{equation}
SP=\hat{\cal S}\prod_{i<j=1}^{n+1}
\Big(\frac{s_{i,j}}{|k_{i\bot}|~|k_{j\bot}|}\Big)^{\alpha_{ij}}
\label{SP}
\end{equation}
instead of $A^{\cal S}_{2\rightarrow
n+2}/{(p_A^+p_B^-)}$. Here the exponents $\alpha_{ij}\sim
\alpha_S$ are different from zero only for some set of
non-overlapping channels and are arbitrary in all other
respects; $\hat{\cal S}$ means symmetrization with
respect to simultaneous change of signs of all $s_{i,j}$
with $i<k\leq j$, performed independently for each
$k=1,\dots,
 n+1$. Indeed, due to the above mentioned
arbitrariness of $\alpha_{ij}$ the fulfilment of the
equalities (\ref{deriv})  for $SP$ guarantees it for any
logarithmic series.

Since we consider only real parts of discontinuities in
the invariants $s_{i,j}$, calculating the discontinuity
of $SP$ in one of $s_{i,j}$ at real  $\alpha_{ij}\sim
\alpha_S$ we can neglect  signs of the other invariants not
only in the leading, but in the NLO, so that we have
\begin{equation}
\label{discontinuites}
    \frac{1}{-\pi i}\left(\sum^{n+1}_{l=j+1}\mathrm{disc}_{s_{j,l}}-
    \sum^{j-1}_{l=0}\mathrm{disc}_{s_{l,j}}\right)SP=
\left(\sum_{l=j+1}^{n+1}\alpha_{jl}
-\sum_{l=0}^{j-1}\alpha_{lj} \right)SP.
\end{equation}
On the other hand, taking into account that
\begin{equation}
\Big(\frac{s_{i,j}}{|k_{i\bot}|~|k_{j\bot}|}\Big)^{\alpha_{ij}}
=e^{\alpha_{ij}(y_i-y_j)},\;\;
\end{equation}
we have for the real part
\begin{equation}
 SP
=e^{\sum_{i<j=1}^{n+1}\alpha_{ij}(y_i-y_j)}\left(1+{\cal
O}(\alpha_S^2)\right),\;\;
\end{equation}
so that, with the NLO accuracy
\begin{equation}\label{derivatives}
\frac{\partial}{\partial y_j}  SP
=\left(\sum_{l=j+1}^{n+1}\alpha_{jl} -\sum_{l=0}^{j-1}\alpha_{lj}
\right)  SP.
\end{equation}
It is clear from Eqs.~(\ref{discontinuites}) and
(\ref{derivatives}) that the  equalities
(\ref{deriv}) are fulfilled.

The important point is that the relations (\ref{deriv})
give a possibility {to find in the NLA  all MRK
amplitudes in all orders of coupling constant}, if they
are known (for all $n$) in the one-loop approximation.
Indeed, these relations express all partial derivatives
of the real parts at some number of loops {through the
discontinuities}, which can be calculated using the
$s$-channel unitarity in terms of amplitudes with a
smaller number of loops; moreover in the NLA only the MRK
is important and only real parts of the amplitudes do
contribute in the unitarity relations. To find ${\cal
A}^{\cal S}_{2\rightarrow n+2}$ besides the derivatives
determined by Eq.~(\ref{deriv}) suitable initial
conditions are required; but since they can be taken at
fixed $y_i$, in the NLA they are necessary only with
one-loop accuracy. Therefore the relations (\ref{deriv})
together with the one-loop approximation for the MRK
amplitudes  unambiguously determine all ${\cal A}^{\cal
S}_{2\rightarrow n+2}$.

Thus,  in order to prove the multi-Regge form (\ref{A
2-2+n}) in the NLA it is sufficient to know that it is
valid in the one-loop approximation and satisfies the equalities
(\ref{deriv}), where the discontinuities are calculated
using this form in the unitarity relations.

Substituting Eq.~(\ref{A 2-2+n}) in the R.H.S. of
Eq.~(\ref{deriv}), we obtain the relations
\begin{equation}
\frac{1}{-\pi
i}\left(\sum_{l=j+1}^{n+1}\mathrm{disc}_{s_{j,l}}
-\sum_{l=0}^{j-1}\mathrm{disc}_{s_{l,j}}\right){\cal
A}^{\cal S}_{2\rightarrow n+2} \, =\,
 \left(\omega(t_{j+1})-\omega(t_{j})\right)
{\cal A}^{R}_{2\rightarrow n+2}~,\label{bootstrap
relations}
\end{equation}
which are called {bootstrap relations}. The
discontinuities in these relations must be calculated
using the $s$--channel unitarity and the multi-Regge form
of the amplitudes (\ref{A 2-2+n}). Evidently, there is an
{infinite number of the bootstrap relations}, because
there is an infinite number of amplitudes ${\cal
A}_{2\rightarrow n+2}$. At the first sight, {it seems a
miracle to satisfy all of them}, since all these
amplitudes are expressed through several Reggeon vertices
and the gluon Regge trajectory. Moreover, {it is quite
nontrivial to satisfy even some definite bootstrap
relation for a definite amplitude}, because it connects
two infinite series in powers of $y_i$, and therefore it
leads to an infinite number of equalities  between
coefficients of these series.

In fact, two miracles must occur in order to satisfy all
the bootstrap relations: first, {for each particular
amplitude ${\cal A}^{R}_{2\rightarrow n+2}$ it must be
possible to reduce the bootstrap relation to a limited
number of restrictions} ({bootstrap conditions}) {on the
gluon trajectory and the Reggeon vertices}, and secondly,
starting from some $n=n_0$ these bootstrap conditions
must be the same as obtained for amplitudes with
$n<n_0$. Finally, {all bootstrap conditions must be
satisfied by the known expressions for the trajectory and the
vertices}.

It is necessary to add  here that  the amplitude in the
R.H.S. of Eq.~(\ref{bootstrap relations}) contains only
colour octets in each of the $q_i$--channel. The
discontinuities in the L.H.S., taken separately, along
with the colour octet hold  other representations of the
colour group, which cancel  in the sum.

\section{Calculation of the discontinuities}

Each of the $s_{i,j}$--channel discontinuity, being
expressed with the help of the $s$--channel unitarity
through the product of amplitudes of the multi-Regge type
(\ref{A 2-2+n}), contains two Reggeons in the channels
$q_l$ at $i<l\leq j$. As an example, the
$s_{j,n+1}$--channel discontinuity is presented
schematically in Fig. 1. Large blobs there stand for
account of the signaturization. In order to present the
discontinuities in a compact way it is convenient to use
operator notations in the transverse momentum and colour
space.   We will use also notations which accumulate all
quantum numbers. Thus, $\langle{\cal G}_i{\cal G}_j|$ and
$|{\cal G}_i{\cal G}_j\rangle$ are $bra$-- and
$ket$--vectors for the $t$--channel states of two
Reggeized gluons with transverse momenta $r_{i \perp}$
and $r_{j \perp}$ and colour indices $c_i$ and $c_j$
correspondingly. It is convenient to distinguish the
states $|{\cal G}_i{\cal G}_j\rangle$ and $|{\cal G}_j
{\cal G}_i\rangle$. We will associate the first of them
with the case when the Reggeon ${\cal G}_i$ is contained
in the amplitude with initial particles (in the lower
part of Fig.1 for the example depicted there), and the
second with the case when it is contained in the
amplitude with final particles (in the upper part of
Fig.1). It is convenient to introduce the scalar product
\begin{equation}\label{norm G}
\langle{\cal G}_i{\cal G}_j|{\cal G}'_{i}{\cal
G}'_{j}\rangle =r^2_{i \perp}r^2_{j \perp}\delta(r_{i
\perp}-r'_{i\perp})\delta(r_{j
\perp}-r'_{j\perp})\delta_{c_ic'_i}\delta_{c_jc'_j}.
\end{equation}
These states are complete, and with the scalar product
(\ref{norm G}) the completeness means
\begin{equation}\label{completeness}
\langle\Psi|\Phi\rangle= \int \frac{d^{D-2}r_{1\bot}
d^{D-2}r_{2\bot}}{r^2_{1\bot}r^2_{2\bot}}\langle\Psi|{\cal
G}_1{\cal G}_2\rangle\langle{\cal G}_1{\cal
G}_2|\Phi\rangle .
\end{equation}
In the following we will also use the letters ${\cal
G}_i$ instead of $c_i$.

%%========================================(FIGURE)===================
% \begin{figure}[h]
% \centering
% %=========================================FIGURE SIMPLE EXAMPLE%
%
% \begin{minipage}{0.85\textwidth}%
% \epsfig{file=new.eps,width=\textwidth}%
% \end{minipage} \\
% \caption{The demonstration of $s_{j,n+1}-$channel discontinuity
% calculation using the rules: the real part of $\frac{1}{-\pi
% i}\mathrm{disc}_{s_{j,n+1}}{\cal A}^{\cal S}_{2\rightarrow n+2} $ is
% represented. Dotted line denotes the discontinuity in the
% corresponding channel.} \label{figexample}
% \end{figure}

\begin{figure}[ht]
\begin{center}
  \begin{picture}(454,245) (6,-15)
    \SetWidth{0.5}
    \SetColor{Black}
    \Vertex(60,52){1.41}
    \SetWidth{2.0}
    \ArrowLine(60,134)(60,201)
    \Line(60,52)(60,134)
    \Text(66,93)[lb]{\huge{\Black{$\dots$}}}
    \Text(52,209)[lb]{\large{\Black{$J_1$}}}
    \Text(91,209)[lb]{\large{\Black{$J_{j-1}$}}}
    \Text(96,24)[lb]{\large{\Black{$\gamma^{J_{j-1}}_{R_{j-1}R_j}$}}}
    \SetWidth{2.5}
    \GOval(150,93)(41,6)(0){0.0}
    \SetWidth{2.0}
    \ArrowLine(150,134)(150,201)
    \Text(101,88)[lb]{\large{\Black{$\langle J_{j} R_{j}|$}}}
    \Text(142,209)[lb]{\large{\Black{$J_{j}$}}}
    \Line(16,52)(16,134)
    \ArrowLine(16,134)(16,201)
    \Text(6,214)[lb]{\Large{\Black{$A'$}}}
    \Line(16,52)(16,11)
    \SetWidth{0.5}
    \Vertex(16,52){2.83}
    \Text(6,-15)[lb]{\Large{\Black{$A$}}}
    \SetWidth{2.0}
    \ArrowLine(330,134)(330,201)
    \Text(322,209)[lb]{\large{\Black{$J_{n}$}}}
    \Text(415,-15)[lb]{\Large{\Black{$B$}}}
    %\Text(278,168)[lb]{\huge{\Black{$\dots$}}}
    %\Text(160,142)[lb]{\large{\Black{$\overbrace{\qquad\qquad\quad\quad}$}}}
    %\Text(341,142)[lb]{\large{\Black{$\overbrace{\qquad\quad\quad\quad}$}}}
    \SetWidth{2.5}
    \GOval(415,93)(41,6)(0){0.0}
    \SetWidth{2.0}
    \Line(415,52)(415,11)
    \ArrowLine(415,129)(415,201)
    \Text(407,209)[lb]{\Large{\Black{$B'$}}}
    \Text(430,88)[lb]{\large{\Black{$|\bar B'B\rangle$}}}
    \SetWidth{1.5}
    \DashLine(114,111)(451,111){10}
    \Text(325,24)[lb]{\large{\Black{$\hat{\cal J}_{n}$}}}
    \SetWidth{2.0}
    \ArrowLine(96,134)(96,201)
    \Line(96,52)(96,134)
    \ArrowLine(250,134)(250,201)
    \Text(245,24)[lb]{\large{\Black{$\hat{\cal J}_{j+1}$}}}
    \Text(235,209)[lb]{\large{\Black{$J_{j+1}$}}}
    \SetWidth{2.5}
    \GOval(250,93)(41,6)(0){0.0}
    \Text(186,88)[lb]{\huge{\Black{$\dots$}}}
    \Text(361,88)[lb]{\huge{\Black{$\dots$}}}
    \Text(278,88)[lb]{\huge{\Black{$\dots$}}}
    \SetWidth{0.5}
    \Vertex(96,52){2.83}
    \SetWidth{1.5}
    \ZigZag(16,52)(250,52){3.5}{26}
    \ZigZag(151,132)(249,132){3.5}{11}
    \ZigZag(330,132)(415,132){3.5}{10}
    %\Text(196,155)[lb]{\large{\Black{$e^{\hat{\cal K}Y_{j+1}}$}}}
    %\Text(370,155)[lb]{\large{\Black{$e^{\hat{\cal K}Y_{n+1}}$}}}
    \SetWidth{0.5}
    \Vertex(60,52){2.83}
    \SetWidth{2.0}
    \ArrowLine(170,52)(170,132)
    \ArrowLine(227,52)(227,132)
    \SetWidth{0.5}
    \Vertex(227,52){2.83}
    \Vertex(170,52){2.83}
    \Vertex(170,132){2.83}
    \Vertex(227,132){2.83}
    \SetWidth{2.5}
    \GOval(330,93)(41,6)(0){0.0}
    \SetWidth{1.5}
    \ZigZag(330,52)(415,52){3.5}{11}
    \ZigZag(250,132)(286,132){3.5}{5}
    \ZigZag(250,52)(281,52){3.5}{4}
    \ZigZag(329,52)(299,52){3.5}{3}
    \ZigZag(329,132)(299,132){3.5}{3}
    \SetWidth{2.0}
    \ArrowLine(271,52)(271,132)
    \ArrowLine(312,52)(312,132)
    \ArrowLine(348,52)(348,132)
    \ArrowLine(394,52)(394,132)
    \SetWidth{0.5}
    \Vertex(271,52){2.83}
    \Vertex(312,52){2.83}
    \Vertex(271,132){2.83}
    \Vertex(312,132){2.83}
    \Vertex(348,132){2.83}
    \Vertex(348,52){2.83}
    \Vertex(394,52){2.83}
    \Vertex(394,132){2.83}
    \Text(47,24)[lb]{\large{\Black{$\gamma^{J_1}_{R_1 R_2}$}}}
    \Text(19,60)[lb]{\large{\Black{$\bar{\Gamma}_{A' A}^{R_1}$}}}
  \end{picture}
\end{center}
\caption{Schematic representation of the
$s_{j,n+1}-$channel discontinuity. The zig-zag lines
depict Reggeized gluon exchanges. The right and left
blobs represent  the $B\rightarrow B'$  and $R_j
\rightarrow J_{j}$ transitions respectively. The
intermediate blobs depict jet productions.}
\label{figexample}
\end{figure}
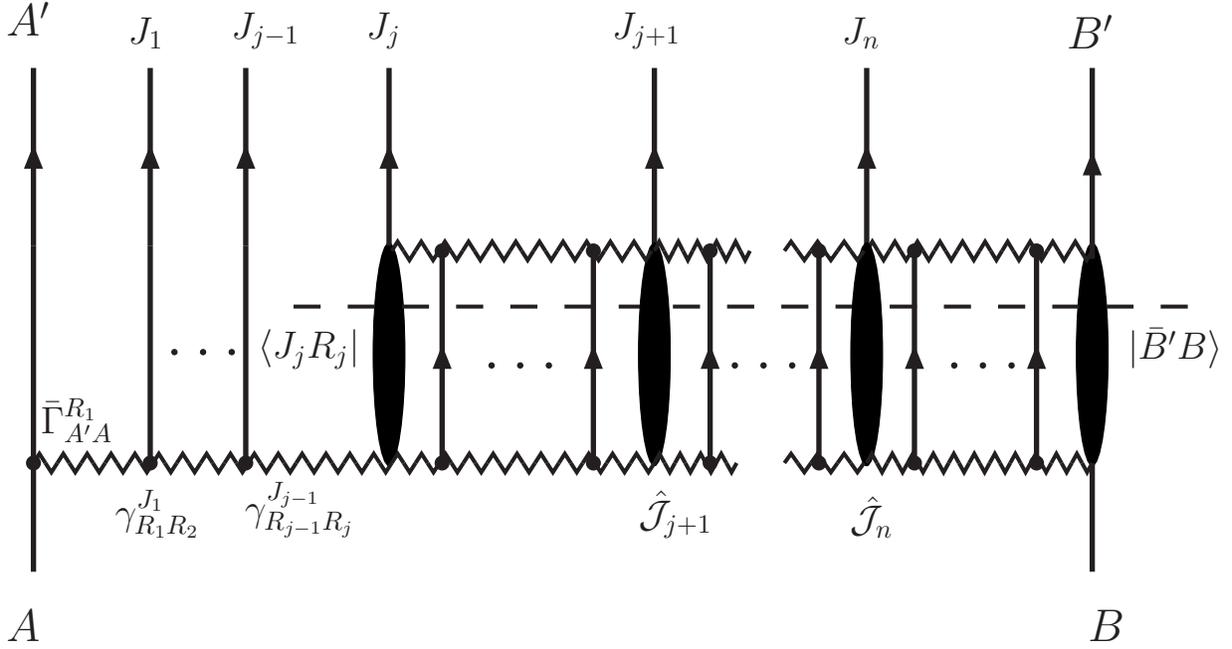

Let us discuss the calculation of the discontinuities.
Our goal is to write them  as  matrix elements of
operator expressions,  consisting of the operators
${\hat{\cal J}}_i$ for jet-$J_i$ production  and of the
operator ${\hat{\cal K}}$ for the Reggeon-Reggeon
interaction kernel, between $bra$-- and $ket$--vectors,
describing either particle-particle or Reggeon-particle
transitions (actually ``particle" here can denote a jet,
as it was already mentioned)  due to interaction with
Reggeized gluons ($R_{j}\rightarrow J_j$ and
$B\rightarrow B'$ transitions in Fig.1). We will call
these states particle-particle or Reggeon-particle
impact-factors.

To calculate the discontinuity we need to convolute
Reggeon vertices with account of the signaturization and
to integrate over momenta of particles in intermediate
states. Since the convolutions of the Reggeon vertices
depend on the transverse components of momenta only, the
signaturization is reduced to anti-symmetrization with
respect to the attached Reggeon lines. In order to escape
double counting in the NLA we introduce an auxiliary
parameter $\Delta\gg 1$ which constrains the difference
in rapidities of particles belonging to one jet.  Note
that the largeness of $\Delta$ is numerical, but not
parametrical (related to $s$), so that terms of order
$\alpha_S\Delta$ are considered as sub-leading. Of
course, the final answer must not depend on $\Delta$.

We denote the momenta of the intermediate jets  $l_\alpha$,
their rapidities $z_\alpha$, with
$z_\alpha=\frac{1}{2}\ln\left(\frac{l^+_\alpha}{l^-_\alpha}\right)$.
Rapidities of the intermediate jets not related neither to
jet-jet or Reggeon-jet transitions, nor to the final jet
production (in the example depicted in  Fig.1 not
contained in the blobs) are confined in the intervals
$[y_{k+1}+\Delta, y_{k}-\Delta]$, where for the
$s_{i,j}$--channel discontinuity $k$ takes values from
$i$ to $j-1$. In each interval we need to perform
integration over rapidities and summation over  number of
jets from $0$ to $\infty$. Denoting $r_{1\perp}$ and
$r_{2\perp}$ the momenta of Reggeons between the jets
$l_\alpha$ and $l_{\alpha+1}$ we write all corresponding
Regge factors in the same form
$e^{(\omega(r_{1\perp})+\omega(r_{2\perp}))(z_\alpha
-z_{\alpha+1})}$. Instead, the Regge factors for the
Reggeons interacting either with the scattering particles
or Reggeons, or with the produced particles (in Fig.1
attached to the blobs) are not uniform. In order to unify
them we include uniformity violating multipliers in the
definitions of jet-production operators and
impact-factors for particle-particle and Reggeon-particle
transitions. After that the two-Reggeon exchange in the
$q_j$-channel is represented by the operator
\begin{equation}\label{series for G}
\hat G (Y_j)^{\Delta}=\sum_{n=0}^\infty
\int^{y_{j-1}-\Delta}_{y_{j}+n\Delta}\!\!\!e^{\hat{\Omega}(y_{j-1}-
z_1)} \hat{\cal K}^{\Delta}_{r} dz_1 \int^{
z_1-\Delta}_{y_{j}+(n-1)\Delta}\!\!dz_2e^{\hat{\Omega}(z_1-z_2)}
\hat{\cal K}^{\Delta}_{r} \dots
\int^{z_{n-1}-\Delta}_{y_{j}+\Delta}dz_n
e^{\hat{\Omega}(z_n-y_{j})}\hat{\cal K}^{\Delta}_{r},
\end{equation}
where the term for $n=0$ is equal to $e^{\hat{\Omega}Y_j}$,
with $\;Y_j=y_{j-1}-y_{j}$, $\hat {\Omega}=\omega(\hat
r_1)+\omega(\hat r_2)$. The operator $\hat{\cal
K}^{\Delta}_r$ takes into account production of
the intermediate jets $J$ with intervals of particle
rapidities $\Delta_J$ in them less than $\Delta$:
\begin{equation}
\langle {\cal G}_1{\cal G}_2|\hat{\cal
K}^{\Delta}_r|{\cal G}'_1{\cal G}'_2\rangle
=\delta(r_{1\bot}+r_{2\bot}-r'_{1\bot}-r'_{2\bot})
\sum_J\int \gamma^J_{{\cal G}_1{\cal
G}'_1}\gamma_J^{{\cal G}_2{\cal G}'_2}
\frac{d\phi_J}{2(2\pi)^{D-1}}\theta(\Delta-\Delta_J).
\label{real kernel}
\end{equation}
Here the sum is taken over all discrete quantum numbers;
$\gamma_J^{{\cal G}_2{\cal G}'_2}$ is the effective
vertex for absorption of  the jet $J$ in the Reggeon
transition ${\cal G}'_2\rightarrow {\cal G}_2$. It is
related to $\gamma^{\bar J}_{{\cal G}_2{\cal G}'_2}$ by
the crossing described above (i.e. by the change of signs of
longitudinal momenta and the corresponding change of wave
functions).  Then, we have
\begin{equation}
d\phi_{J}=\frac{d k^2_J}{2\pi} (2\pi)^{D}
\delta^D(k_J-\sum_i p_i)
\prod_{i}\frac{d^{D-1}p_{i}}{\left( 2\pi
\right)^{D-1}2\epsilon _{i}}~ \label{phi jet}
\end{equation}
for a jet $J$ with total momentum $k_J$ consisting of
particles with momenta $p_i$. The integration limits in
Eq.~(\ref{series for G}) correspond to the limitation on the
intervals of particle rapidities in Eq.~(\ref{real kernel}).

As it was already mentioned, terms of order
$\alpha_S\Delta$ are sub-leading, therefore we need to
retain in Eq.~(\ref{series for G}) only terms linear in $\Delta$
with coefficients of order $\alpha_S$. With this
accuracy we can write $\hat G
(Y_j)^{\Delta}=\bigl(1-\hat{\cal K}_r^B\Delta\bigr)\hat G
(Y_j)\bigl(1-\hat{\cal K}_r^B\Delta\bigr)$; the
superscript $B$ here and below denotes leading order, so
that ${\cal K}_r^B$ is given by ${\cal O}(\alpha_S)$
terms in Eq.~(\ref{real kernel}), and $\hat G (Y_j)$ is
obtained from Eq.~(\ref{series for G}) by the omission of
$\Delta$ in the integration limit and the replacement
$\hat{\cal K}^{\Delta}_{r}\rightarrow \hat{\cal K}_{r}$, where
\begin{equation}
  \hat{\cal K}_r=\hat{\cal
K}^{\Delta}_r-\hat{\cal K}^{B}_r\hat{\cal K}^{B}_r
\Delta. \label{subtracted real kernel}
\end{equation}
We include the multipliers $\bigl(1-\hat{\cal
K}_r^B\Delta\bigr)$ in definitions of jet-production
operators and impact-factors for particle-particle and
Reggeon-particle transitions. Then the two-Reggeon
exchange in the $q_j$-channel is represented by the
operator $\hat G (Y_j)$. It is easy to see that it obeys
the equation $d\hat G (Y)/dY =\hat{\cal K}\hat G (Y)$,
where \be\hat{\cal K}=\omega(\hat r_1)+\omega(\hat
r_2)+\hat{\cal K}_r. \label{kernel}\ee Using the initial
condition $\hat G (0)=1$ we obtain
\begin{equation}
\hat G (Y) = e^{\hat{\cal K}Y}.  \label{Green exponent}
\end{equation}

With account of the terms discussed before
Eq.~(\ref{series for G}) and after Eq.~(\ref{subtracted
real kernel}) the impact-factor for the $B\rightarrow B'$
transition is defined as
\begin{equation}
|\bar B^\prime B\rangle=|\bar B^\prime
B\rangle^{\Delta}-\left({\omega^B(\hat
r_1)}\ln\left|\frac{\hat r_{1\bot}}{q_{B\bot}}\right|
+{\omega^B(\hat r_2)}\ln\left|\frac{\hat
r_{2\bot}}{q_{B\bot}}\right|
 + \hat{\cal K}^B_r\;\Delta\right)|\bar B^\prime
B\rangle^{B},
\label{BB}
\end{equation}
where
\begin{equation}
\langle {\cal G}_1{\cal G}_2|\bar B^\prime
B\rangle^{\Delta} = \delta(q_{B\bot}-r_{1\bot}-r_{2\bot})
\frac{1}{2p_B^-}\sum_{ \tilde B  }\int
\left(\Gamma^{{\cal G}_1}_{\tilde B B} \Gamma^{{\cal
G}_2}_{B' \tilde B }-\Gamma^{{\cal G}_2}_{\tilde B B}
\Gamma^{{\cal G}_1}_{B' \tilde B }\right)
 d\phi_{ \tilde B}\prod_l\theta(\Delta -(z_l-y_B))~.
 \label{BBdelta}
\end{equation}
Here $q_{B}=p_B-p_{B^\prime}$ and $z_l$ are the
rapidities of particles in intermediate jets. The terms
with $\omega^B$ in Eq.~(\ref{BB}) takes into account the
difference of the Regge factors related to the Reggeons
interacting with the particles $B$ and $B'$ and the
``uniform" factors used in the series (\ref{series for
G}) for $\hat G (Y_{n+1})^{\Delta}$. The term with
$\hat{\cal K}^B_r$ in Eq.~(\ref{BB}) comes from the relation
between $\hat G (Y_{n+1})^{\Delta}$ and $\hat G
(Y_{n+1})$. Note that in the case when  $B$ or $B'$ is a
two-particle jet, only the first term must be kept in
Eq.~(\ref{BB}); moreover, only the Born approximation for
this term  must be taken in  Eq.~(\ref{BBdelta}).

It is clear that for the impact-factor of the
$A\rightarrow A'$ transition we have
\begin{equation}
\langle A^\prime \bar A|=\langle A^\prime \bar
A|^\Delta-\langle A^\prime \bar A|^B\left({\omega^B(\hat
r_1)}\ln\left|\frac{\hat r_{1\bot}}{q_{A\bot}}\right|
+{\omega^B(\hat r_2)}\ln\left|\frac{\hat
r_{2\bot}}{q_{A\bot}}\right|
 + \hat{\cal K}_r^B\;\Delta\right),
 \label{AA}
\end{equation}
\begin{equation}
\langle A^\prime \bar A|{\cal G}_1{\cal
G}_2\rangle^{\Delta} =
\delta(q_{A\bot}-r_{1\bot}-r_{2\bot})
\frac{1}{2p_A^+}\sum_{ \tilde A }\int\left( \Gamma^{{\cal
G}_1}_{\tilde A A } \Gamma^{{\cal G}_2}_{A' \tilde A
}-\Gamma^{{\cal G}_2}_{\tilde A A } \Gamma^{{\cal
G}_1}_{A' \tilde A }\right)
 d\phi_{ \tilde A}\prod_l\theta(\Delta -(y_A-z_l))~,
 \label{AAdelta}
\end{equation}
where $q_{A}=p_{A^\prime}-p_A$.

The anti-symmetrization with respect to the permutation
${\cal G}_1\leftrightarrow{\cal G}_2$ in Eqs.~(\ref{BBdelta})
and (\ref{AAdelta}) takes into account  the signaturization. The
important fact is that due to the signaturization only the
antisymmetric colour octet survives from all possible
colour states of the two Reggeons ${\cal G}_1$ and ${\cal
G}_2$. For quark and gluon impact-factors it follows from
results of Ref.~\cite{FFKPIF}. For the case when some
state is a two-particle  it can be seen from results
presented in Ref.~\cite{VF03}.

Accordingly, the Reggeon-particle impact-factors are
defined as
\[
 |\bar J_i R_{i+1}\rangle=|\bar J_i R_{i+1}\rangle^\Delta
 -\left(\left({\omega(q_{i+1})}-{\omega(\hat
r_1)}\right)\ln\left|\frac{k_{i\bot}}
{(q_{(i+1)\bot}-\hat r_{1\bot})}\right| \right.
\]
\begin{equation}
\left. -{\omega(\hat r_2)}\ln\left|\frac{k_{i\bot}}{\hat
r_{2\bot}}\right|
 + \hat{\cal K}_r^{B}\Delta\right)|\bar J_i
 R_{i+1}\rangle^B,
\end{equation}
\[
\langle {\cal G}_{1} {\cal G}_{2}|\bar J_i
R_{i+1}\rangle^\Delta =
\delta(q_{(i+1)\bot}-k_{i\bot}-r_{1\bot}-r_{2\bot})
\]
\begin{equation}
\times\frac{1}{2k^-_J} \sum_{ J}\int \left(\gamma^{J
}_{{\cal G}_1 R_{i+1}} \Gamma^{{\cal G}_2}_{J_i J
}-\gamma^{J }_{{\cal G}_2R_{i+1}} \Gamma^{{\cal
G}_1}_{J_i J }\right)
 d\phi_{ J }\prod_{l}\theta(\Delta-(z_l-y_i)),
\end{equation}\\
and
\[
\langle J_i  R_i|=\langle J_i R_i|^\Delta-\langle J_i
 R_i|^ B\left(\left({\omega(q_i)}-\omega(\hat
r_1)\right)\ln\left|\frac{k_{i\bot}} {(q_{i\bot}-\hat
r_{1\bot})}\right|\right.
\]
\begin{equation}
\left.-{\omega(\hat r_2)}\ln\left|\frac{k_{i\bot}}{\hat
r_{2\bot}}\right|
 + \hat{\cal K}_r^{B}\Delta\right),
\end{equation}
\[
\langle J_i  R_i|{\cal G}_{1}{\cal G}_{2}\rangle^{\Delta}
= \delta(r_{1\bot}+r_{2\bot}-q_{i\bot}-k_{i\bot})
\]
\begin{equation}
\times\frac{1}{2k_J^+}\sum_{ J}\int \left(\gamma^{J
}_{R_i{\cal G}_1} \Gamma^{{\cal G}_2}_{J_i J }
 -\gamma^{J
}_{R_i{\cal G}_2} \Gamma^{{\cal G}_1}_{J_i J }\right)
d\phi_{ J }\prod_{l}\theta(\Delta -(y_i-z_l)).
\end{equation}

At last, the  operators ${\hat{\cal J}}_i$ for production
of jets $J_i$ are defined as
\[
\hat{\cal J}_i=\hat{\cal J}_i^{\Delta}-\left( \hat{\cal
K}^{B}_r\hat{\cal J}^{B}_i+\hat{\cal J}^{B}_i\hat{\cal
K}^{B}_r\right)\;\Delta, \;\;\; \langle {\cal
G}^\prime_{1}{\cal G}^\prime_{2}|\hat{\cal \cal
J}^{\Delta}_i|{\cal G}_{1}{\cal G}_{2}\rangle=
\]
\[=\delta(r_{1\perp}+r_{2\perp}-k_{i\perp}-r^\prime_{1\perp}-
r^\prime_{2\perp})  \left[ \gamma^{J_i}_{{\cal G}_1{\cal
G}^\prime_1}\delta(r_{2\perp}-r^\prime_{2\perp})
 r_{2\perp}^{~2}\delta_{{\cal
G}_2{\cal G}_2^\prime}+\gamma^{J_i}_{{\cal G}_2{\cal
G}^\prime_2}\delta(r_{1\perp}-r^\prime_{1\perp})
 r_{1\perp}^{~2}\delta_{{\cal
G}_1{\cal G}_1^\prime}\right.
\]
\begin{equation}
\left.+\sum_{G }\int_{y_i-\Delta}^{y_i+\Delta}
\frac{dz_{G}}{2(2\pi)^{D-1}}\left(\gamma^{\{ J_i
G\}}_{{\cal G}_1{\cal G}^{\prime}_1}\gamma^{{\cal
G}_2{\cal G}^\prime_2}_{G}+ \gamma^{G}_{{\cal G}_1{\cal
G}^\prime_1}\gamma^{{\cal G}_2{\cal G}^\prime_2}_{J_i G}
\right)\right]. \label{jet-delta}
\end{equation}
Here the last term appears only in the case when
$J_i\equiv G_i$ is a single gluon,  the sum in this term
goes over quantum numbers of the intermediate gluon $G$
and the vertices  must be taken  in the Born
approximation. At that $\gamma^{\{ J_i G\}}_{{\cal
G}_1{\cal G}^{\prime}_1}$ is the vertex for production of
the jet consisting of the gluons $G_i$ and $G$,
$\;\;\gamma^{{\cal G}_2{\cal G}^\prime_2}_{G_i G}$ is the
vertex for absorption of the gluon $G$ and production of
the gluon $G_i$ at the ${\cal G}_2\rightarrow {\cal
G}^\prime_2$ transition; it can be obtained from
$\gamma_{{\cal G}_2{\cal G}^\prime_2}^{\{G_i G\}}$ by
crossing with respect to the gluon $G$.

With the definitions given above we obtain
\[
-4i(2\pi)^{D-2}\delta(q_{(j+1)\bot}-q_{i\bot}-\sum_{l=i}^{l=j}
k_{l\bot})\;\mathrm{disc}_{s_{i,j}}A^{\cal S}_{2\rightarrow
n+2}=\bar{\Gamma}^{ R_1}_{
A^{\prime}A}\frac{e^{\omega(q_1)(y_{0}-y_1)}}{q^2_{1\perp}}\times
\]
\begin{equation}\label{disc s ij}
\times \left(
\prod_{l=2}^{i}\gamma^{J_{l-1}}_{{R}_{l-1}
 {R}_{l}}
 \frac{e^{\omega(q_l)(y_{l-1}-y_l)}}{q^2_{l\perp}}
 \right)
 \langle J_i
R_i|\left( \prod_{l=i+1}^{j-1}
 e^{\hat{\cal K}(y_{l-1}-y_l)}\hat{\cal J}_l \right)
 e^{\hat{\cal K}(y_{j-1}-y_j)} |\bar J_j R_{j+1}\rangle\times
\end{equation}
\[
\times \left(
\prod_{l=j+1}^{n}
 \frac{e^{\omega(q_l)(y_{l-1}-y_l)}}{q^2_{l\perp}}\gamma^{J_l}_{{R}_l
 {R}_{l+1}}
 \right)
 \frac{e^{\omega(q_{n+1})(y_{n}-y_{n+1})}}{q^2_{(n+1)\perp}}
\Gamma^{R_{n+1}}_{ B^{\prime} B}.
\]

If  $i=0$ we must omit all factors  on the left from
$\langle J_0  R_0|$ and substitute $\langle J_0 R_0|$
with  $\langle A^{\prime} \bar A |,
\;\;q_0+k_0$ with $p_{A'}-p_A$; in the case $j=n+1$
we must omit all factors on the right from  $ |\bar
J_{n+1} R_{n+2}\rangle$ and perform the substitutions
$|\bar J_{n+1} R_{n+2}\rangle \rightarrow |\bar
B^{\prime} B \rangle, \;\; q_{n+2}-k_{n+1}\rightarrow
p_B-p_{B'}$.

\section{Bootstrap conditions}

Let us prove, using the representation (\ref{disc s ij})
for the discontinuities,   that an infinite number of the
bootstrap relations (\ref{bootstrap relations}) are
satisfied if the following bootstrap conditions are
fulfilled: the impact-factors for scattering particles
satisfy equations \be |\bar B' B\rangle =
{g}\Gamma^{R_{n+1}}_{B'B}|R_{\omega}(q_{B\perp})\rangle,
\;\; \langle{{A'\bar A}}|
={g}\bar{\Gamma}^{R_1}_{A'A}\langle
R_{\omega}(q_{A\perp})|, \label{second bootstrap} \ee
where $\langle R_{\omega}(q_{\perp})|$ and
$|R_{\omega}(q_{\perp})\rangle $   are the $bra$-- and
$ket$-- vectors of the universal (process
independent) eigenstate of the kernel $\hat {\cal K}$
with the eigenvalue $\omega(q_{\perp})$, \be
 \hat {\cal K}|R_{\omega}(q_{\perp})\rangle = \omega(q_{\perp})
 |R_{\omega}(q_{\perp})\rangle, \;\;\langle R_{\omega}(q_{\perp})|
 \hat {\cal K} = \langle R_{\omega}(q_{\perp})|\,\omega(q_{\perp})
,\label{first bootstrap} \ee and the normalization is fixed
through the scalar product \be
\frac{g^2q^2_{\perp}}{2(2\pi)^{D-1}}\langle
R_{\omega}(q'_{\perp})|R_{\omega}(q_{\perp})\rangle=-
\delta(q_{\perp}-q'_{\perp})\omega(q_{\perp})~;
\label{normalization}\ee the Reggeon-gluon impact-factors
and the gluon production vertices satisfy the equations
\[
\hat{\cal{J}}_{i}\,|R_{\omega}(q_{(i+1)\bot})\ra\: g\:
q^2_{(i+1)\bot}+|\bar J_i R_{i+1}\ra=
|R_{\omega}(q_{i\bot})\ra\,g\:\gamma^{J_i}_{R_{i}R_{i+1}}
\,,
\]
\begin{equation}\label{third bootstrap}
g\:q^2_{i\bot} \langle
R_{\omega}(q_{i\bot})|\hat{\cal{J}}_{i}\, +\langle J_i
 R_{i}|= g\:\gamma^{J_i}_{R_{i}R_{i+1}}\langle
R_{\omega}(q_{(i+1)\bot})| \,,
\end{equation}
where $q_{(i+1)\bot}=q_{i\bot}+k_{i\bot}$. Actually the
second of Eqs.~(\ref{second bootstrap}), (\ref{first
bootstrap}) and (\ref{third bootstrap}) are not independent
since  $bra$-- and $ket$--vectors are related with each
other by the change of $+$ and $-$ momenta components.

The bootstrap conditions  (\ref{second bootstrap}) and
(\ref{first bootstrap}) are known since a long time
\cite{FF98}-\cite{FFKP00} and have been proved to be
satisfied \cite{FFKPIF}-\cite{FP02}.  The bootstrap
relations for {elastic amplitudes} require only a weak
form of the conditions (\ref{second bootstrap}) and
(\ref{first bootstrap}), namely only the projection of
these conditions  on $|R_{\omega}\rangle$. It was
recognized \cite{V.F.02} that the bootstrap relations for
{one-gluon production amplitudes}  besides the conditions
(\ref{second bootstrap}) and (\ref{first bootstrap})
require also a weak form of the condition (\ref{third
bootstrap}). Thus, {the bootstrap relations for one-gluon
production amplitudes play a twofold role: they
strengthen the conditions imposed by the elastic
bootstrap and give a new one}. One could expect that the
history will repeat itself upon addition of each next
gluon in the final state. If it were so, we would have to
consider the bootstrap relations for production of an
arbitrary number of gluons and would obtain an infinite
number of bootstrap conditions. Fortunately, the  history
is repeated only partly: it was shown \cite{BFF03} that
already {the bootstrap relations for two-gluon production
require the strong form of the last condition (i.e.
Eq.~(\ref{third bootstrap})) and do not require new
conditions}.

The bootstrap conditions with  two-particle jets are
required  in the NLA  only with the Reggeon vertices
taken in the Born approximation. They were checked and
proved to be satisfied in Refs.~\cite{FKR03} and \cite{VF03}. After
that only the condition (\ref{third bootstrap}) remained not evident.
Its fulfilment was proved recently \cite{FKR-tbp}. Thus,
now it is shown that all bootstrap conditions are
fulfilled.

To prove that the bootstrap conditions (\ref{second
bootstrap})--(\ref{third bootstrap}) secure the fulfilment of
all infinite set of the bootstrap relations
(\ref{bootstrap relations}),  consider first the terms
with $l=n$ and $l=n+1$ in the representation
(\ref{bootstrap relations}). Using this last representation for
the discontinuities  and applying the bootstrap
conditions (\ref{second bootstrap}) and (\ref{first
bootstrap}) to the $s_{k,n+1}$--channel discontinuity, we
obtain that the sum of the discontinuities in the
channels $s_{k,n}$ and $s_{k,n+1}$ contains
\begin{equation}\label{bootstrap proof 1}
g\:\hat{\cal{J}}_{n}\,|R_{\omega}(q_{(n+1)\bot})\ra +|\bar J_n
\!R_{n+1}\ra\frac{1}{q^2_{(n+1)\perp}}=
|R_{\omega}(q_{n\bot})\ra\,g\:
\gamma^{J_n}_{R_{n}R_{n+1}}\frac{1}{q^2_{(n+1)\perp}}\,.
\end{equation}
The equality here follows from the bootstrap condition
(\ref{third bootstrap}). Now the procedure can be
repeated: we can apply to this sum  the bootstrap
condition (\ref{first bootstrap}), and to the sum of the
obtained result with the $s_{k,n-1}$--channel
discontinuity  again Eq.~(\ref{third bootstrap}). Thus
all sum over $l$ from $j+1$ to $n+1$ in the
representation (\ref{bootstrap relations}) is reduced to
one term. A quite analogous procedure (with the use of the
bootstrap conditions for $bra$--vectors) can be applied
to the sum over $l$ from $0$ to $j-1$. As a result we
have that the left part of the representation
(\ref{bootstrap relations}) with the coefficient $
-2(2\pi)^{D-1}\delta(q_{(j+1)\bot}-q_{j\bot}-k_{j\bot})$,
where $q_{(j+1)\bot}=p_{B\bot}-p_{B'\bot}
-\sum_{l=j+1}^{l=n} k_{l\bot}$ and $
q_{j\bot}=p_{A'\bot}-p_{A\bot} +\sum_{l=1}^{l=j-1}
k_{l\bot}$, can be  obtained from the R.H.S. of the
multi-Regge form (\ref{A 2-2+n}) by the replacement
\begin{equation}
\gamma^{J_j}_{R_jR_{j+1}}\longrightarrow\; \langle J_j
R_j|R_\omega (q_{(j+1)\bot})\rangle g
q^2_{(j+1)\bot}-gq^2_{j\bot}\langle R_\omega
(q_{j\bot})|\bar J_j R_{j+1}\rangle.
\end{equation}
Taking the difference of the first equality in
the condition (\ref{third bootstrap}) for $i=j$ multiplied by
$gq^2_{j\bot}\langle R_\omega
(q_{j\bot})|$ and the second equality multiplied by $|
R_\omega (q_{(j+1)\bot})\rangle g q^2_{(j+1)\bot}$ and
using the normalization (\ref{normalization}) we obtain
\[
 \langle J_j R_j|R_\omega
(q_{j+1})\rangle\,g\,q^2_{(j+1)\bot}-g\,q^2_{j\bot}
\,\langle R_\omega (q_{j})|\bar J_j R_{j+1}\rangle
=-2(2\pi)^{D-1}\delta(q_{(j+1)\bot}-q_{j\bot}-k_{j\bot})
\]
\begin{equation}
\times \left(\omega(q_{j+1})-
\omega(q_{j})\right)\gamma^{J_j}_{R_jR_{j+1}} .
\end{equation}
That concludes the proof.

Thus, the fulfilment of the bootstrap conditions (\ref{second
bootstrap})--(\ref{third bootstrap}) guarantees
the implementation of all the infinite set of the bootstrap
relations (\ref{bootstrap relations}).

\section{Summary}
We presented the basic steps of the proof that in the
multi--Regge kinematics real parts of QCD amplitudes for
processes with gluon exchanges have the simple
multi-Regge form depicted in Eq.~(\ref{A 2-2+n}), with
the accuracy up to next-to-leading logarithms. This
statement is extremely powerful. An infinite number of
QCD processes is described by several Reggeon vertices
and the gluon Regge trajectory.  This remarkable property
of QCD amplitudes is extremely important for the
description of high energy processes. In particular, it
appears as the basis of the BFKL approach.

The proof is based on the bootstrap relations required by
the compatibility of the multi--Regge form (\ref{A
2-2+n}) of inelastic QCD amplitudes with the $s$--channel
unitarity. It consists of several steps.  First, we
proved in Section 3 that the multi--Regge form (\ref{A
2-2+n}) is guaranteed in all orders of perturbation
theory if it is valid in the one-loop approximation and
if the set of the bootstrap relations (\ref{bootstrap
relations}) holds true. These relations contain the
$s$--channel discontinuities of inelastic amplitudes
which must be calculated using the unitarity relations
and  the multi--Regge form (\ref{A 2-2+n}). Then, to find
a representation for the discontinuities we developed the
operator formalism introduced in Ref.~\cite{FFKP00} for
taking into consideration inelastic amplitudes. This
permitted us to find in Section 4 the closed expressions
(\ref{disc s ij}) for the discontinuities in terms of the
Reggeon vertices and the gluon Regge trajectory. The last
step, performed in Section 5, concerns the proof that the
bootstrap relations (\ref{bootstrap relations}) are
fulfilled if the vertices and trajectories submit to the
bootstrap conditions (\ref{second
bootstrap})--(\ref{third bootstrap}).  It is extremely
nontrivial that  an infinite set of the bootstrap
relations is reduced to several conditions on the Reggeon
vertices and the gluon Regge trajectory. All these
vertices, as well as the gluon Regge trajectory are known
now in the next-to-leading order.  The bootstrap
conditions  were examined for a long time in a series of
papers with increase of understanding of their role (see
for instance \cite{FP02} and references therein). On the
parton level (for quarks and gluons) only the condition
(\ref{third bootstrap}) remained unchecked till recently.
Now the fulfilment of this  condition is proved
\cite{FKR-tbp}.

To be  rigorous we have to say that strictly speaking the
form (\ref{A 2-2+n}) in the one-loop approximation was
actually derived only for one-gluon production
\cite{Fadin:1993wh}.  Although there are general
arguments that it is correct for any $n$, a strong
evidence is absent.

\vspace{18pt}
\goodbreak
\noindent
{\bf Acknowledgements}
\vskip 6pt
V.S. Fadin thanks the Dipartimento di Fisica dell'Universit\`a
della Calabria and the I\-sti\-tu\-to Nazionale di Fisica Nucleare -
gruppo collegato di Cosenza for their warm hospitality while a
part of this work was done.

%\newpage


\begin{thebibliography}{99}
%1.......................................................
\bibitem{GST}
M.T. Grisaru, H.J. Schnitzer and H.-S. Tsao,  Phys. Rev.
Lett. {\bf 30}, 811 (1973); Phys. Rev. D {\bf 8}, 4498
(1973).
%2.......................................................
\bibitem{L76}
L.N. Lipatov, Yad. Fiz. {\bf 23}, 642 (1976) [Sov. J.
Nucl. Phys. {\bf 23}, 338 (1976)].
%3.......................................................
\bibitem{BFKL}
V.S. Fadin, E.A. Kuraev and L.N. Lipatov, Phys. Lett. B
{\bf 60}, 50 (1975); E.A. Kuraev, L.N. Lipatov and V.S.
Fadin, Zh. Eksp. Teor. Fiz. {\bf 71}, 840 (1976) [Sov.
Phys. JETP {\bf 44}, 443 (1976)]; Zh. Eksp. Teor. Fiz.
{\bf 72}, 377 (1977) [Sov. Phys. JETP {\bf 45}, 199
(1977)]; Ya.Ya. Balitskii and L.N. Lipatov, Yad. Fiz.
{\bf 28}, 1597 (1978) [Sov. J. Nucl. Phys. {\bf 28}, 822
(1978)].
%4.......................................................
\bibitem{FS}
V.S. Fadin and V.E. Sherman, Pis'ma  Zh. Eksp. Teor. Fiz.
{\bf 23}, 599 (1976) [Sov. Phys. JETP Lett. {\bf 23}, 548
(1976)]; Zh. Eksp. Teor. Fiz. {\bf 72}, 1640 (1977) [Sov.
Phys. JETP {\bf 45}, 861 (1977)]; V.S.~Fadin and
R.~Fiore,
%``Calculation of Reggeon vertices in QCD,''
Phys.\ Rev.\ D {\bf 64}, 114012 (2001).
%5.......................................................
\bibitem{BDFG}
A.V.~Bogdan, V.~Del Duca, V.S.~Fadin and E.W.~Glover,
%``The quark Regge trajectory at two loops,''
JHEP {\bf 0203}, 032 (2002).
%6.......................................................
\bibitem{KLPV}
M.I.~Kotsky, L.N.~Lipatov, A.~Principe and
M.I.~Vyazovsky,
%``Radiative corrections to the quark-gluon-Reggeized quark vertex
%in QCD,''
Nucl. Phys. {\bf B 648}, 277 (2003).
%7.......................................................
\bibitem{BLF}
Ya.Ya.~Balitskii, L.N.~Lipatov and V.S.~Fadin, in {\it
Materials of IV Winter School of LNPI} (Leningrad, 1979)
p.109.
%8.......................................................
\bibitem{Lipatov:1995pn}
  L.~N.~Lipatov,
  %``Gauge invariant effective action for high-energy processes in QCD,''
  Nucl.\ Phys.\ B {\bf 452} (1995) 369;
  %``Small-x physics in perturbative QCD,''
  Phys.\ Rept.\  {\bf 286} (1997) 131.
%9.......................................................
\bibitem{FL98} V.S. Fadin and L.N. Lipatov, Phys. Lett. B429 (1998)
127; M. Ciafaloni and G. Camici, Phys. Lett. B430 (1998)
349.
%10.......................................................
\bibitem{FF05}
V.~S.~Fadin and R.~Fiore,  Phys.\ Lett.\ B {\bf 610}
(2005) 61 [Erratum-ibid.\ B {\bf 621} (2005) 61];  Phys.\
Rev.\ D {\bf 72} (2005) 014018.
%11.......................................................
\bibitem{Antonov:2004hh}
  E.~N.~Antonov, L.~N.~Lipatov, E.~A.~Kuraev and I.~O.~Cherednikov,
  %``Feynman rules for effective Regge action,''
  Nucl.\ Phys.\ B {\bf 721} (2005) 111.
%12.......................................................
\bibitem{Kucs}
 T. Kucs, arXiv:hep-ph/0403023.
%13.......................................................
\bibitem{V.F.02}
V.S.~Fadin,
%``Justification of the BFKL approach in the NLA,''
{\it Talk given at the NATO Advanced Research Workshop
"Diffraction 2002",  August 31 - September 6, 2002,
Alushta, Crimea, Ukraine},  in {\it Diffraction 2002},
Ed. by R. Fiore {\it et al.}, NATO Science Series, Vol.
101, p.235.
%14.......................................................
\bibitem{VF03}
V.~S.~Fadin,
%``Multi-Reggeon processes in QCD,"
Phys.\ Atom.\ Nucl.\  {\bf 66} 2017  (2003).
%15-----------------------
\bibitem{Bartels_1}
J. Bartels, Phys. Rev. \textbf{D11} 2977, 2989 (1975);
Nucl. Phys. \textbf{B175}, 365 (1980).
%16.......................................................
\bibitem{S:1960}
O. Steinmann, Helv. Phys. Acta, \textbf{33}, 33 (1960).
%17.......................................................
\bibitem{FFKPIF}
V.S.~Fadin, R.~Fiore, M.I.~Kotsky and A.~Papa,
%``The Gluon Impact Factors,''
Phys.\ Rev.\ D {\bf 61}, 094005 (2000);
%``The quark impact factors,''
{\bf 61}, 094006 (2000).
%18.......................................................
\bibitem{FF98}
V.S.~Fadin and R.~Fiore, Phys. Lett. B {\bf 440}, 359
(1998).
%19.......................................................
\bibitem{Braun99}
M.~Braun and G.P.~Vacca, Phys. Lett. B {\bf 454}, 319;
M.~Braun, hep-ph/9901447 (1999).
%20.......................................................
\bibitem{FFKP00}
V.S.~Fadin, R.~Fiore, M.I.~Kotsky and A.~Papa,
%``Strong Bootstrap Conditions,''
Phys. Lett. B {\bf 495}, 329 (2000).
%21.......................................................
\bibitem{BV00}
M.~Braun and G.P.~Vacca, Phys. Lett. B {\bf 477}, 156
(2000).
%22.......................................................
\bibitem{FFP99}
V.S.~Fadin, R.~Fiore and A.~Papa,
%``The Quark Part of the Non-forwardBFKL Kernel and the 'Bootstrap'for the
%Gluon Reggeization,''
Phys. Rev. D {\bf 60}, 074025 (1999);  V.S.~Fadin,
R.~Fiore and M.I.~Kotsky,
%``The compatibility of the gluon Reggeization with the s-channel unitarity,''
Phys. Lett. B {\bf 494}, 100 (2000).
%23.......................................................
\bibitem{FP02}
V.~S.~Fadin and A.~Papa,
%``A proof of fulfillment of the strong bootstrap condition,''
Nucl.\ Phys.\ B {\bf 640}, 309 (2002).
%24.......................................................
\bibitem{BFF03}
J.~Bartels, V.~S.~Fadin and R.~Fiore,
%``The bootstrap conditions for the gluon reggeization,''
Nucl.\ Phys.\ B {\bf 672} 329 (2003).
%25.......................................................
\bibitem{FKR03} V.S.~Fadin, M.G.~Kozlov and A.V.~Reznichenko,
%``Radiative corrections to QCD amplitudes in quasi-multi-Regge  kinematics,''
Yad.\ Fiz.\ {\bf 67} (2004) 377[Phys.\ Atom.\ Nucl.\ {\bf
67}  359 (2004)].
%26......................................................
\bibitem{FKR-tbp} V.S.~Fadin, M.G.~Kozlov and A.V.~Reznichenko,
to be published.
%27......................................................
\bibitem{Fadin:1993wh} V.~S.~Fadin and L.~N.~Lipatov,
  %``Radiative corrections to QCD scattering amplitudes in a multi - Regge
  %kinematics,''
  Nucl.\ Phys.\ B {\bf 406} (1993) 259.


\end{thebibliography}
\end{document}